\documentclass[%
reprint,
 amsmath,amssymb,
prb,
]{revtex4-1}

\usepackage{physics}% load physics package
\usepackage{graphicx}% Include figure files
\usepackage{dcolumn}% Align table columns on decimal point
\usepackage{bm}% bold math
\usepackage{hyperref}% add hypertext capabilities
\usepackage{cleveref}
\usepackage[utf8]{inputenc}
\begin{document}

%\preprint{APS/123-QED}

\title{Phase Diagram Reconstruction of the Bose-Hubbard Model with a Restricted Boltzmann Machine Wavefunction}% Force line breaks with \\

\author{Vladimir Vargas-Calderón}
\email{vvargasc@unal.edu.co}
\author{Herbert Vinck-Posada}%
\affiliation{%
 Grupo de Superconductividad y Nanotecnología, Departamento de Física, Universidad Nacional de Colombia, Bogotá, Colombia
}%


\author{Fabio A. González}
\affiliation{
 Computing Systems and Industrial Eng. Department, Universidad Nacional de Colombia, Bogotá, Colombia
}%

\date{\today}% It is always \today, today,

\begin{abstract}
Recently, the use of neural quantum states for describing the ground state of many- and few-body problems has been gaining popularity because of their high expressivity and ability to handle intractably large Hilbert spaces. In particular, methods based on variational Monte Carlo have proven to be successful in describing the physics of bosonic systems such as the Bose-Hubbard (BH) model. However, this technique has not been systematically tested on the parameter space of the BH model, particularly at the boundary between the Mott insulator and superfluid phases. In this work, we evaluate the capabilities of variational Monte Carlo with a trial wavefunction given by a Restricted Boltzmann Machine to reproduce the quantum ground state of the BH model on several points of its parameter space. To benchmark the technique, we compare its results to the ground state found through exact diagonalization for small one-dimensional chains. In general, we find that the learned ground state correctly estimates many observables, reproducing to a high degree the phase diagram for the first Mott lobe and part of the second one. However, we find that the technique is challenged whenever the system transitions between excitation manifolds, as the ground state is not learned correctly at these boundaries. We improve the quality of the results produced by the technique by proposing a method to discard noisy probabilities learned in the ground state.
\end{abstract}

\maketitle

%\tableofcontents
\section{Introduction}
The dimension of a Hilbert space that describes the possible states of a many-body system scales exponentially with the number of one-body states and with the number of particles. In most cases, this feature comes as an important practical difficulty for physicists to study many-body problems, because approximate techniques have to be implemented in order to perform simulations (e.g., dynamical mean-field theory~\citep{georges1996dmft}, density matrix renormalization group (DMRG)~\citep{schollwock2005dmrg}). One of the most studied quantum objects is the ground state of a many-body system for a number of reasons: particles tend to occupy the lowest energy states first, as dictated by the aufbau principle; also excited states inherit the ground state structure; and most prominently, it is an object that encodes changes in observables that exhibit quantum phase transitions. Recently, a technique to calculate the ground state of a many-body system has been proposed, based on defining a trial wavefunction with a neural network, which is optimized to minimize the ground state energy through variational Monte Carlo(VMC)~\citep{Carleo2017science}. This technique has proven to be successful in approximating with high fidelity the ground state of several condensed matter many- and few-body systems. Some examples are atomic and molecular systems~\citep{Han2019,choo2019fermionic,pfau2019abinitio,hermann2019deep,kessler2019artificial,Ruggeri2018}, the transverse-field Ising model~\citep{carleo2018constructing,sharir2020deepautoregressive,hibatallah2020recurrent} with quenching~\citep{Czischek2018}, the Heisenberg model~\citep{carleo2018constructing} and its anti-ferromagnetic version~\citep{sharir2020deepautoregressive,Nomura2017}, the quantum harmonic oscillator in electric field~\citep{teng2018radial}, the Hubbard model~\citep{Nomura2017}, the $J_1$-$J_2$ Heisenberg model~\citep{hibatallah2020recurrent} and its antiferromagnetic version~\citep{szab2020neural,Choo2019twodim,Liang2018solving}, as well as models for interacting bosons in one and three-dimensions~\citep{saito2018method}.

The success in approximating the ground state comes from two sources. The first one is technical: neural networks have expressivity and are able to approximate arbitrary functions~\citep{Gao2017}, and their optimization methods in machine learning have been thoroughly studied~\citep{sun2019survey}. The second one is physical: even though the size of a Hilbert space exponentially grows with respect to one-body states and number of particles, only a small set of those states are needed to describe the ground state~\citep{Carleo2017science}. Here the definition of small can vary, as it will be seen in this work.

Although reported results indicate that the quantum ground state can be represented through neural network wavefunctions, most of these studies focus on the convergence of the energy. Only a few of them use the technique to characterize quantum correlations near a quantum phase transition to study the phase space of a Hamiltonian comprehensively. Indeed, the convergence of the energy comes quicker than the convergence of the state, as any first-order error in the variational state leads to only a second-order error in its corresponding energy~\citep{ballentine1998quantum}. Therefore, focusing on energy convergence might be misleading when asserting the power of VMC with neural network trial wavefunctions. In contrast to previous works, we intensively and systematically test the capabilities of VMC to reconstruct, through a Restricted Boltzmann Machine (RBM) wavefunction, the quantum ground state of a one-dimensional BH system throughout the superfluid (SF) and Mott insulator (MI) phases. We chose the BH model because of the known numerical difficulties near the MI-SF boundary. Consequently, we contrast exact diagonalization solutions with the ones provided by the VMC-RBM method in systems of 5 and 8 sites for a fine mesh in the BH parameter space at the first and part of the second Mott lobes. It is only due to the sweeping of the BH parameters that we are able to see that points near the MI-SF boundary are challenging for the VMC-RBM method. However, we also find that some of the differences between the learned ground state and the ground state found through exact diagonalization originate from badly learned probability amplitudes of Fock states that should have small contribution to the ground state. In this work, we also propose a state cleaning technique that removes those badly learned probabilities.

There have also been previous efforts to learn the ground state of the BH Hamiltonian through different neural network wavefunctions. However, as mentioned before, these studies do not extensively test the VMC framework throughout the BH parameter space. For instance, a permutation symmetric RBM implemented in NetKet~\cite{carleo2019netket} has been used to study the energy and particle density convergence at two points in the SF and MI phases, finding a difficulty in the convergence to the numerically exact particle density in the SF phase~\cite{mcbrian2019ground}. However, a qualitatively good location of the boundary between the SF and MI phases for the region enclosing the first two Mott lobes was achieved using the particle density as a discriminator~\cite{mcbrian2019ground}. RBMs have also been used to show how for a fixed number of bosons in the system, the learned ground state is able to replicate the numerically exact order parameter of the quantum phase transition~\cite{liquantum}. However, fixing the number of bosons greatly reduces the size of the Hilbert space, making it easier for the VMC technique to learn the ground state since a more complete sampling of the Hilbert's basis states is possible. Also, the BH model has been considered as a toy model for trying full-forward neural networks (FFNNs) and convolutional neural networks (CNNs) wavefunctions. For instance, the ground state of one and two-dimensional finite lattices with parabolic confinement potential for a fixed number of bosons was approximated through an FFNN ansatz~\citep{Saito2017}. In the case of no confinement and periodic boundary conditions (which introduce displacement symmetry), both FFNNs and RBMs were used in large 1D lattices of up to 40 sites~\citep{Choo2018}. CNNs have also been introduced and compared with FFNN wavefunctions to approximate this ground state~\citep{Saito2018}. A CNN is also proposed as a means to introduce the ground state at finite temperature in a 1D lattice for larger on-site boson interaction than hopping interaction~\citep{Irikura2019}. Despite the important results derived from these works, they do not focus on the ability of the neural network ansatz to reproduce the ground state near the quantum phase transition boundary. Nevertheless, unsupervised learning has been previously used to classify states as belonging to the Mott insulator or SF phases in the BH model with previously obtained states or physical quantities, for several values of the order parameters~\cite{Liu2018,Huembeli2018,broecker2017,perez2019detection}.

We believe that the ability of VMC technique to reproduce the ground state near the quantum phase transition has to be extensively tested. Therefore, we pay close attention to the description of the variational ground state near the SF-MI phase for a small number of sites to compare it with the numerically exact ground state. This paper is organized as follows. \Cref{sec:model} exposes the main features of the BH Hamiltonian and describes the SF-MI transition in the system. In this section, an overview of the VMC technique is given, discussing the difficulties that must be faced and overcome in order to learn the ground state of the BH model. In \cref{sec:results}, the main results are given, including energy convergence, the overlap of the learned and exact ground states, the reproduction of the phase diagram via two different order parameters, as well as tomographies indicating relevant~\footnote{By relevant, we mean that a Fock state $\ket{\vb*{n}}$ has a large probability amplitude with respect to other Fock states.} Fock states for the ground state. Finally, in \cref{sec:conclusion} conclusions of this work are given.

\section{Model and Variational Monte Carlo\label{sec:model}}

\subsection{Bose-Hubbard Hamiltonian}
The BH Hamiltonian describes the interactions between bosons that can occupy sites in a $d$-dimensional lattice. The model is characterized by a hopping energy $t$, an on-site interaction $U$ and a chemical potential $\mu$, so that the grand-canonical Hamiltonian reads ($\hbar=1$)~\citep{Fisher1989}
\begin{align}
    \hat{H} = -t\sum_{\langle ij \rangle} (\hat{a}_i^\dagger \hat{a}_j + \text{H.c.}) + \frac{U}{2}\sum_i \hat{n}_i(\hat{n}_i-1) - \mu \sum_i \hat{n}_i,\label{eq:bhHam}
\end{align}
where $\hat{a}_i$ is the annihilation operator at site $i$, and $\hat{n}_i = \hat{a}^\dagger_i \hat{a}_i$ is the number operator. The notation $\langle ij \rangle$ indicates that the sum runs over pairs of neighbor sites in the lattice. The BH model is able to reproduce experimental results in Josephson-junction networks~\citep{VanOudenaarden1996,Baltin1997,bruder2005bose} and in lattices of ultra-cold atoms~\citep{Greiner2003,Jaksch1998,Bloch2008,stoeferle2004,spielman2007,gemelke2009situ}. The latter offers precise control of the lattice parameters~\citep{Grynberg2000,kollath2004,lugan2007}. Theoretically, a lot of attention has been devoted both to understand the quantum phases of the system (the ones that arise from \cref{eq:bhHam} and from the disordered or extended BH model with longer range interactions)~\citep{Freericks1996,Ejima2012,Cazalilla2011,Teichmann2009,batrouni1995,krauth1991,Krauth1991iop,niyaz1991}. Another central issue is to calculate the quantum phase transition boundaries~\citep{Kuhner1998,Ejima2011,Kuhner2000,batrouni1992world,Kashurnikov1996,batrouni1990critical,elstner1999,Koller2006}, whose precision has improved over the years with better calculation techniques and computing power, revealing features such as the re-entrance phenomenon~\citep{Pino2013,elstner1999}, where for particular values of the chemical potential, the system switches between the MI phase to the SF phase, and back to the MI phase before definitely entering the SF phase after an increase of $t/U$.

For simplicity, we will restrict our analyses to the $d=1$ case, where only two quantum phases are possible. When the on-site interaction energy is much larger than the hopping energy, the latter becomes negligible, and the Hamiltonian is written as the sum of independent Hamiltonians for each site $\frac{U}{2} \hat{n}_i (\hat{n}_i - 1) - \mu\hat{n}_i$. These one-particle Hamiltonians can be immediately diagonalized by the number basis. The corresponding eigenenergies are $\frac{U}{2}n_i(n_i-1) - \mu n_i$, which reach minimum values for fixed $\mu$ and $U$ at $n_i = \max\{0, \lceil \mu/U \rceil\}$ (note that all sites are equivalent). Moreover, in this regime, the expected variance of the local number operator is 0, i.e. $\langle \hat{n}_i^2\rangle - \langle \hat{n}_i\rangle^2 = 0$. This regime characterizes the MI phase. On the other hand, when the hopping energy is much larger than the on-site interaction energy, the latter becomes negligible. Thus, the Hamiltonian can also be written as the sum of independent Hamiltonians, but in momentum space, where $\tilde{a}_k = N^{-1/2}\sum_{j=1}^N\hat{a}_j e^{-ix_jp_k/\hbar}$ is the boson annihilation operator in the momentum representation. Here, $x_j=c\times j$, where $c$ is the lattice constant, and $p_k = 2\pi k \hbar / (N\times c)$. Each independent Hamiltonian in momentum space ($\sum_k(-2t\cos(2\pi k/N)-\mu)\tilde{a}^\dagger_k\tilde{a}_k$) has eigenenergies $-2t\cos(2\pi k/N) - \mu$, which reach their minimum when all bosons condense with 0 momenta. Note that the energies are independent of the states' occupation, meaning that the ground state is degenerate for any number of particles. This regime is known as the SF phase, characterized by a delocalized wavefunction (formally described by algebraic decaying spatial correlations~\citep{Ejima2012,Pino2013}, which is why the SF phase in 1D is not a true Bose-Einstein condensate). Thus, the ground state has a non-zero expected variance of the local number operator. In fact, the probability distribution for the local occupation is Poissonian, meaning that $\expval*{n_i^2} - \expval*{n_i}^2 = \expval*{n_i}$~\citep{Greiner2003}. For a fixed chemical potential and an infinite number of sites, there exists a continuous phase transition from the MI phase to the SF phase as $U$ decreases with respect to $t$ (with the exception of a range of $\mu$ values in the first Mott lobe, where re-entrance exists).

\subsection{Variational Monte Carlo with Restricted Boltzmann Machine Wave Function}

\begin{figure}[!ht]
    \centering
    \includegraphics[width=0.7\columnwidth]{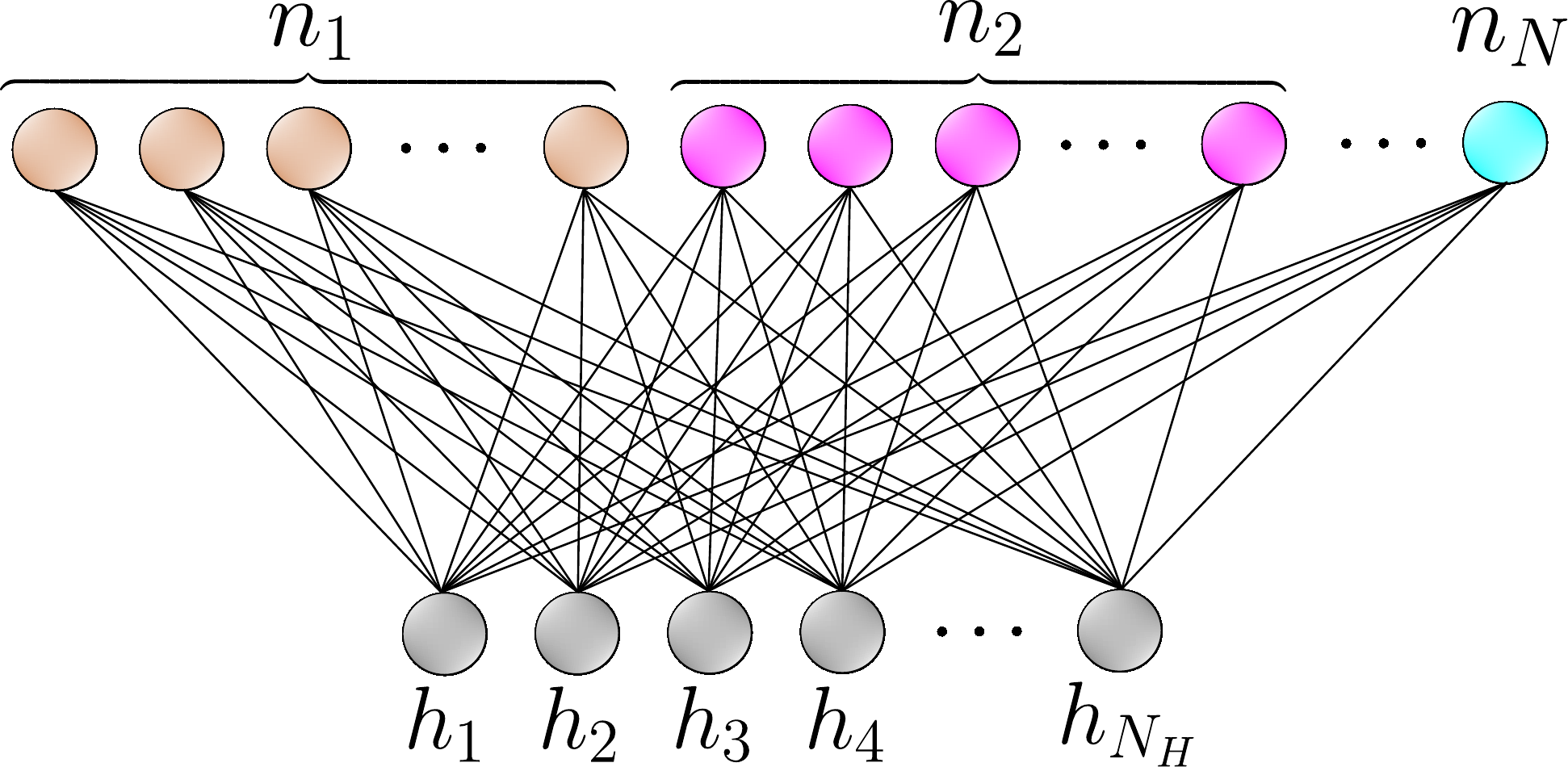}
    \caption{(Color online) Illustration of the RBM, where each site occupation is one-hot encoded into $M$ visible neurons, depicted with different colors for different sites in the top layer. There are $N_H$ hidden neurons in the bottom layer connected with the visible neurons through weights $W_{\ell,j}$ that connect the $\ell$-th hidden neuron with the $j$-th visibile neuron.}
    \label{fig:rbm}
\end{figure}

In this section, a review of the method introduced by~\citet{Carleo2017science} is given, where a VMC setup is used to find the ground state, formally written as a trial wavefunction given by an RBM. In the Fock space basis, the ground state of the BH model can be written as~\cite{Saito2017}
\begin{align}
    \ket{\Psi} = \sum_{n_1=0,n_2=0,\ldots}^\infty \Psi(n_1,n_2,\ldots) \ket{n_1,n_2,\ldots},
\end{align}
where $n_i$ is the occupation number of the $i$-th site, and $|\Psi(n_1,n_2,\ldots)|^2$ are the probability amplitudes corresponding to the Fock states $\ket{n_1,n_2,\ldots}$. In order to map the wave function to a computer, both the number of sites and the number of possible particles in each site have to be truncated. We will refer to the number of sites as $N$ and to the maximum number of particles in each site as $M-1$. The coefficients $\Psi(n_1,n_2,\ldots,n_N)$ are approximated by an RBM. RBMs are generative neural networks, formally described by a bipartite undirected graph such as the one shown in \cref{fig:rbm}, where there is a layer of visible neurons denoted by $\vb*{v}$ that are used to input real data, and a layer of hidden neurons denoted by $\vb*{h}$ that are used as latent variables of the model~\citep{smolensky1986}. In particular, the wavefunction coefficients take the form $ \Psi(n_1,n_2,\ldots,n_N) \approx \psi_{\vb*{\theta}}(n_1,\ldots,n_N) = \sum_{\vb*{h}} e^{-E_{\text{RBM}}(\vb*{v}(\vb*{n}), \vb*{h})}$, where $E_{\text{RBM}}(\vb*{v}(\vb*{n}), \vb*{h})$ is the energy of the RBM~\citep{smolensky1986}, and $\vb*{\theta}$ are the variational parameters of the RBM. As a short-hand notation, an occupation configuration is denoted as $\vb*{n}$, and it is inputted to the visible layer of the RBM. However, the configuration first needs to be one-hot encoded as follows: each occupation $n_i$ is encoded into an $M$-component vector whose $j$-th component is $\delta_{j,n_i}$, $j=0,1,\ldots, M-1$; then, the vectors for every occupation are concatenated into $\vb*{v}(\vb*{n})$. Moreover, if the $N_H$ hidden neurons are restricted to binary values $1$ or $-1$, then, the approximated wavefunction coefficients can be written as~\cite{Carleo2017science}
\begin{align}
      \psi_{\vb*{\theta}}(\vb*{n})=e^{\sum_{j} a_jv_j(\vb*{n})}\prod_{\ell=1}^{N_H}2\cosh\left(b_\ell + \sum_j W_{\ell,j}v_j(\vb*{n})\right),
\end{align}
where $a_j, b_\ell$ and $W$ are the complex-valued visible bias, hidden bias and connection matrix of the RBM, respectively.

\begin{figure}[!ht]
    \centering
    \includegraphics[width=0.6\columnwidth]{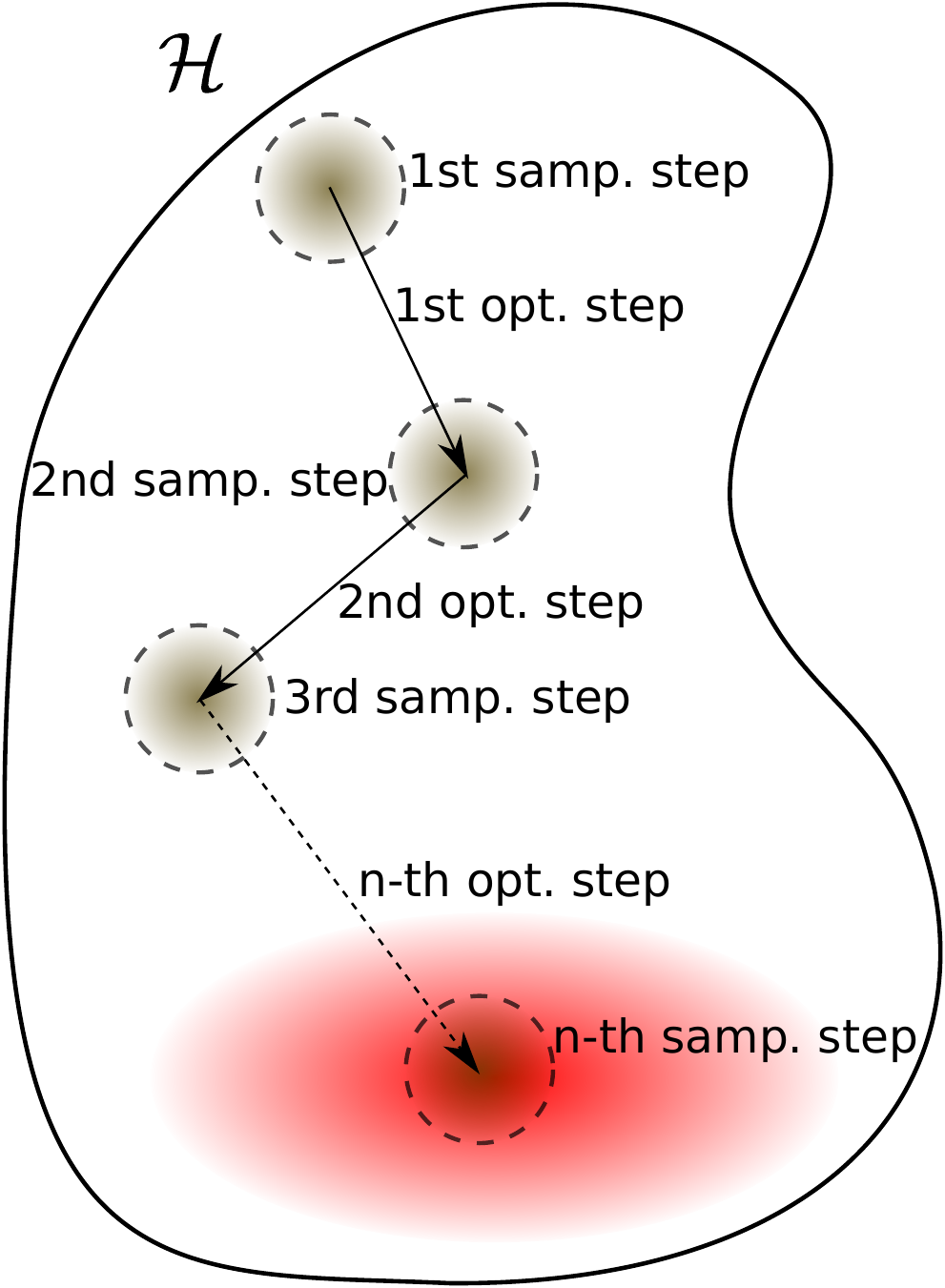}
    \caption{(Color online) Representation of the variational Monte Carlo technique. With randomly initialized parameters $\vb*{\theta}$, a set of states from the Hilbert space $\mathcal{H}$ is sampled. By minimizing the energy defined in \cref{eq:energy}, the parameters $\vb*{\theta}$ are updated. These two steps are repeated $n$ times with the objective of sampling the states (in the occupation basis) that are relevant for the ground state, depicted by a blot at the bottom, with the correct probability distribution.}
    \label{fig:vmc}
\end{figure}

Thus, the approximation of the wavefunction coefficients is done through the adjustment of the parameters $\vb*{\theta}:\{a_j,b_\ell, W_{\ell,j}\}$ that minimize the energy $\bra{\psi_{\vb*{\theta}}}\hat{H}\ket{\psi_{\vb*{\theta}}}$. At each step of the minimization, a set $\mathcal{M}$ of configurations $\vb*{n}$ is sampled from $|\psi_{\vb*{\theta}}(\vb*{n})|^2$ using the Metropolis-Hastings algorithm, so that the energy can be efficiently estimated as~\citep{Saito2017}
\begin{align}
    \bra{\psi_{\vb*{\theta}}}\hat{H}\ket{\psi_{\vb*{\theta}}} \approx \frac{1}{|\mathcal{M}|}\sum_{\vb*{n}\in\mathcal{M}}\sum_{\vb*{n}'}\bra{\vb*{n}}\hat{H}\ket{\vb*{n}'}\frac{\psi_{\vb*{\theta}}(\vb*{n}')}{\psi_{\vb*{\theta}}(\vb*{n})}.\label{eq:energy}
\end{align}
More explicitly, the following steps are carried out to generate the sample $\mathcal{M}$. At the first iteration, a state $\vb*{n}_0$ is randomly proposed. Then, at the $i$-th iteration:
\begin{enumerate}
    \item Under some updating rule, propose a new state $\vb*{n}'_i$ from $\vb*{n}_i$.
    \item With probability $\min\{1, |\psi_{\vb*{\theta}}(\vb*{n}'_i) / \psi_{\vb*{\theta}}(\vb*{n}_i)|^2\}$ accept the state $\vb*{n}'_i$, i.e. $\vb*{n}_{i+1} \leftarrow \vb*{n}'_i$. If it is not accepted, then $\vb*{n}_{i+1} \leftarrow \vb*{n}_i$.
\end{enumerate}
To build the sample $\mathcal{M}$ a total of 1000 iterations are performed. The updating rule is provided by NetKet's Local Metropolis sampler, which picks a site at random and assigns to it a uniformly random occupation, except from the current occupation. Then, through either stochastic gradient descent or stochastic reconfiguration~\citep{sorella2007}, the energy in \cref{eq:energy} can be minimized, producing a new set of parameters $\vb*{\theta}$, as explained by \citet{Carleo2017science}. Both the sampling and minimization of the energy with respect to the RBM parameters are repeated iteratively until the RBM parameters converge, resembling an Expectation Maximization algorithm~\citep{dempster1977EMalgorithm}. A schematic representation of the variational Monte Carlo technique is shown in~\cref{fig:vmc}.

The blot at the bottom of~\cref{fig:vmc} depicts a region of the Hilbert space that contains the most relevant Fock states $\ket{\vb*{n}}$ for the ground state of the BH Hamiltonian, with fixed $\mu, t$ and $U$. For instance, the only relevant Fock state within the $m$-th Mott lobe is $\ket{n_1=m, n_2=m,\ldots}$, whereas in the SF phase there are many relevant Fock states. In that regard, the region of relevant Fock states can contain a variable number of Fock states depending on the parameters of the Hamiltonian that one is intending to solve. Therefore, an issue immediately arises: for an unknown target probability distribution $|\Psi(\vb*{n})|^2$, the sampling can be too small or too large with respect to the size of the region of relevant Fock states. If it is small, important information about the ground state might not be taken into account, whereas if it is large, noisy probability from non-relevant states can be taken into account.

\section{Results\label{sec:results}}

We swept several values of chemical potential and hopping energy (a grid of $72\times100$ points) corresponding to the first and part of the second Mott lobe in the $t/U$--$\mu/U$ space, with $U=1$, performing 12000 sampling and optimization steps with NetKet~\citep{carleo2019netket}, where 1000 Metropolis-Hastings steps were done for each sampling step. This was done for three different scenarios: for 5 sites, we used 8 and 20 hidden neurons, and for 8 sites, we used 11 hidden neurons. In all cases, the maximum number of bosons allowed per site was 4.

\subsection{Energy convergence}

\begin{figure}[!ht]
    \centering
    \includegraphics[width=\columnwidth]{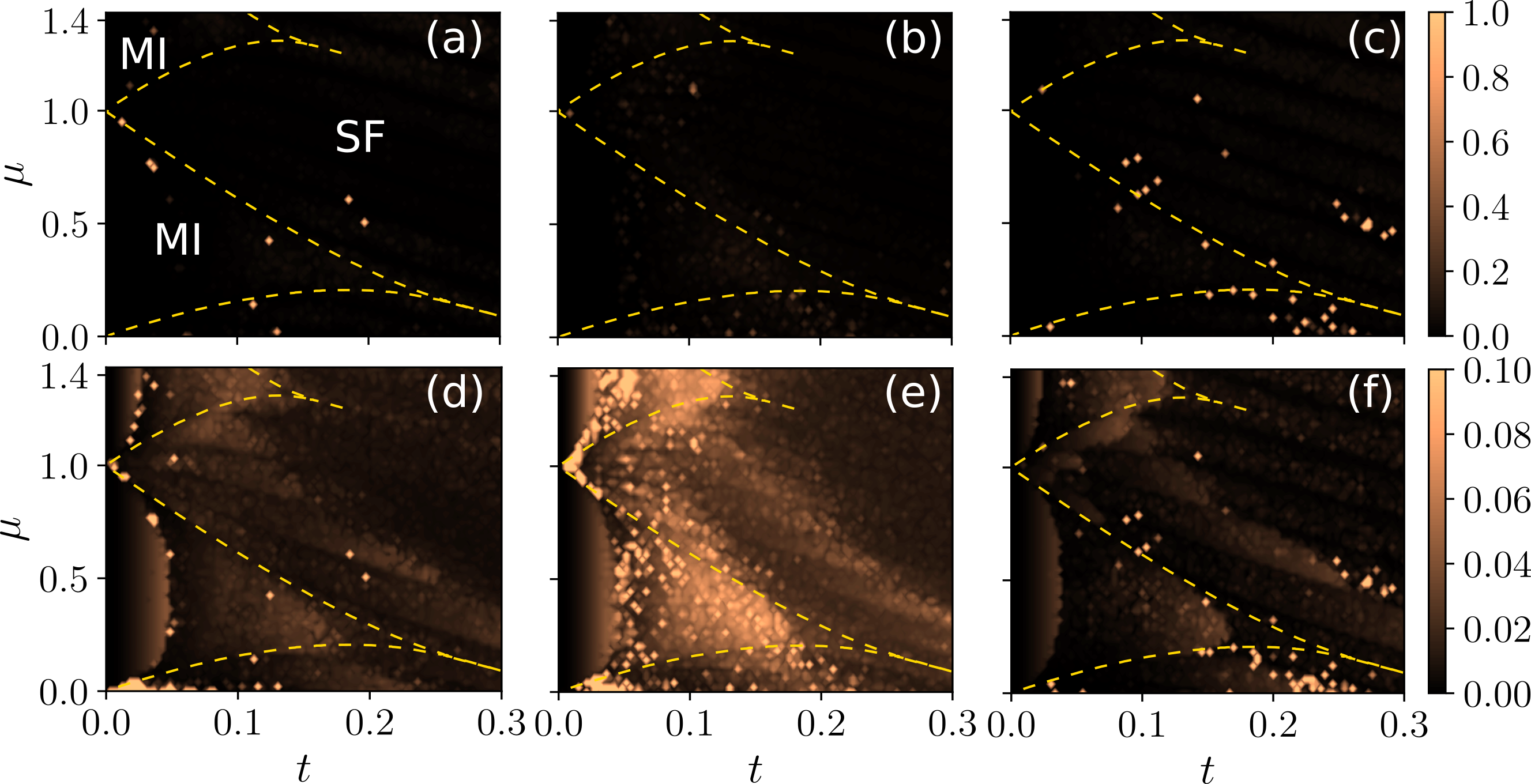}
    \caption{(Color online) Energy variance and absolute error of the last 500 sampling-optimization steps. Dashed lines show the phase boundaries computed for 128 sites with DMRG by \citet{Ejima2011}. In (a), the two first Mott lobes and the SF region are labeled explicitly. (a)-(c) show the variance for the last 500 sampling-optimization steps for the cases of 5 sites and 8 hidden neurons, 8 sites and 11 hidden neurons, and 5 sites and 20 hidden neurons, respectively. (d)-(f) show the corresponding absolute errors between the average energy value for the last 500 sampling-optimization steps and the exact ground state energy.}
    \label{fig:energy_convergence}
\end{figure}

Since the VMC minimizes the energy, we check that the energy computed with \cref{eq:energy} converges by measuring its variance for the last 500 sampling-optimization steps, as well as by measuring the absolute error of the energy when compared to the exact ground state energy found through Lanczsos diagonalization. For the aforementioned three scenarios, the variance for the last 500 sampling-optimization steps is shown in \cref{fig:energy_convergence}(a)-(c), and the corresponding absolute errors with respect to the exact ground state energy are shown in \cref{fig:energy_convergence}(d)-(f). It is seen that in the Hamiltonian parameter space, the majority of the energies have low-variance, showing convergence towards a value that is in excellent agreement with the exact ground-state energies.

\begin{figure}[!ht]
    \centering
    \includegraphics[width=\columnwidth]{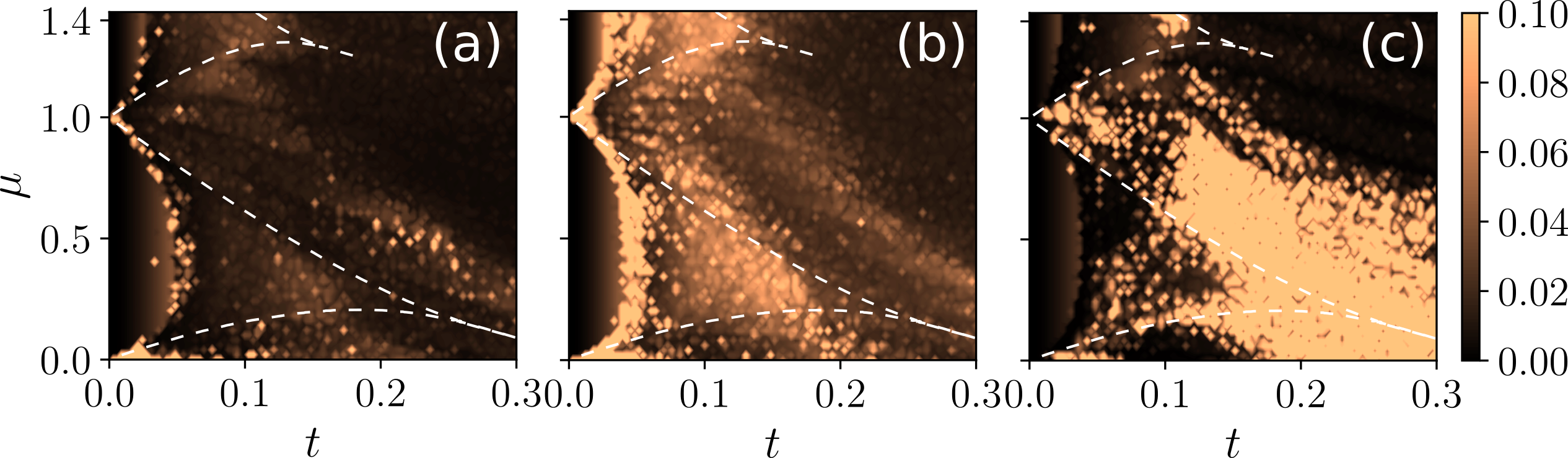}
    \caption{(Color online) Absolute error between the RBM state expected energy computed through \cref{eq:exactrbm} and the exact ground state energy for (a) 5 sites and 8 hidden neurons, (b) 8 sites and 11 hidden neurons and (c) 5 sites and 20 hidden neurons. Dashed lines are the MI-SF boundaries as in \cref{fig:energy_convergence}.}
    \label{fig:mae_rbm_energy}
\end{figure}

Since the Hilbert spaces are small enough to compute all the probability amplitude coefficients for every state in the Fock basis, we can also compute any observable $\hat{O}$ as
\begin{align}
    \bra{\psi_{\vb*{\theta}}}\hat{O}\ket{\psi_{\vb*{\theta}}} = \frac{\sum_{\vb*{n}\in\mathcal{H}} |\psi_{\vb*{\theta}}(\vb*{n})|^2 \bra{\vb*{n}}\hat{O}\ket{\vb*{n}}}{\sum_{\vb*{n}\in\mathcal{H}} |\psi_{\vb*{\theta}}(\vb*{n})|^2 }. \label{eq:exactrbm}
\end{align}
In particular, the absolute error of the energy computed through \cref{eq:exactrbm} is shown in \cref{fig:mae_rbm_energy}. It is clear that in the case of 5 sites and 20 neurons, a very large region both in the SF and MI phases shows a strong disagreement between the energy calculated through \cref{eq:energy,eq:exactrbm}, showing that even though the energy converged, the state did not. Another recurrent pattern in the energies calculated through \cref{eq:energy,eq:exactrbm} is an arc of high absolute errors formed in the left-most side of the Mott lobes, which we will address later.

\subsection{Overlap}
\begin{figure}[!ht]
    \centering
    \includegraphics[width=\columnwidth]{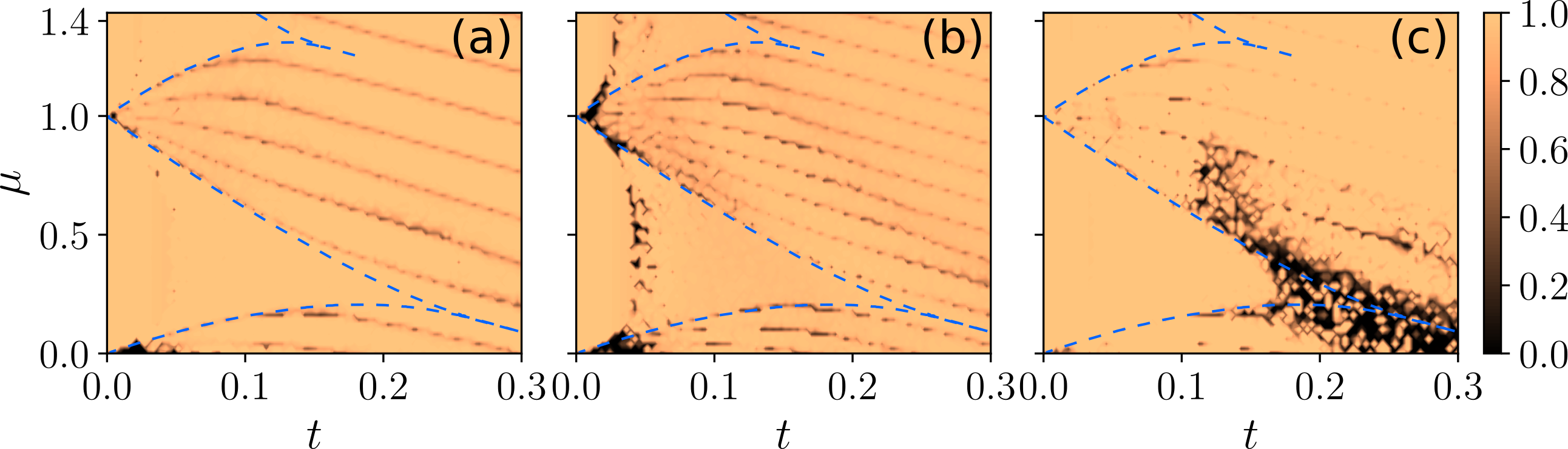}
    \caption{(Color online) Overlap $\abs{\braket{\psi_\text{exact}}{\psi_{\vb*{\theta}}}}^2$ between the exact and RBM ground states for (a) 5 sites and 8 hidden neurons, (b) 8 sites and 11 neurons, and (c) 5 sites and 20 neurons. Dashed lines are the MI-SF boundaries as in \cref{fig:energy_convergence}.}
    \label{fig:overlapRBMexact}
\end{figure}

The lack of convergence of the RBM state for the case of 5 sites and 20 hidden neurons is further confirmed when we measure the overlap between the exact ground state $\ket{\psi_{\text{exact}}}$ and the RBM ground state $\ket{\psi_{\vb*{\theta}}}$, shown in \cref{fig:overlapRBMexact}(c). It is now clear why the expected energy with respect to the complete RBM state shown in \cref{fig:mae_rbm_energy}(c) presents large errors when compared to the expected energy with respect to the exact ground state: it appears that the RBM has not learned the ground state in the bottom-right region of the plot, which is a region that covers part of the Mott lobe, as well as part of the SF phase region. Moreover, in the case of 5 sites and 8 hidden neurons, and the case of 8 sites and 11 hidden neurons, the RBM finds difficulty in learning the ground state in the limit between the MI and the SF phase as shown in \cref{fig:overlapRBMexact}(a) and (b). The difficulty in learning those states, and in general, in treating the ground state near the MI-SF boundary comes from the Kosterlitz-Thouless-like quantum phase transition in 1D systems~\citep{giamarchi2003quantum}, where an exponentially small Mott gap exists~\citep{Ejima2012}. We see once again that there are arcs of low overlap points formed in the left-most side of the Mott lobes. Within the SF phase, there are also lines of low overlap, which appear because of finite size effects. Note that there are as many of these fictitious boundaries as there are sites in the periodic chain under study. Nevertheless, when comparing the 5 sites cases, it is seen that for values of $\mu/U > 1$ the states at the MI-SF phases boundary are better learned when 20 hidden neurons are used in the RBM (as in \citep{mcbrian2019ground}, cf. \cref{fig:overlapRBMexact}(c)) than when only 8 hidden neurons are used (see \cref{fig:overlapRBMexact}(a)).

An interesting motif is that if one pays very close attention to the energy absolute errors in \cref{fig:energy_convergence}(d)-(f), the lowest errors occur where the overlap in \cref{fig:overlapRBMexact}(a)-(c) is lowest. Therefore, even though the convergence in the energy is excellent, the state is not learned correctly. 

\begin{figure}[!ht]
    \centering
    \includegraphics[width=\columnwidth]{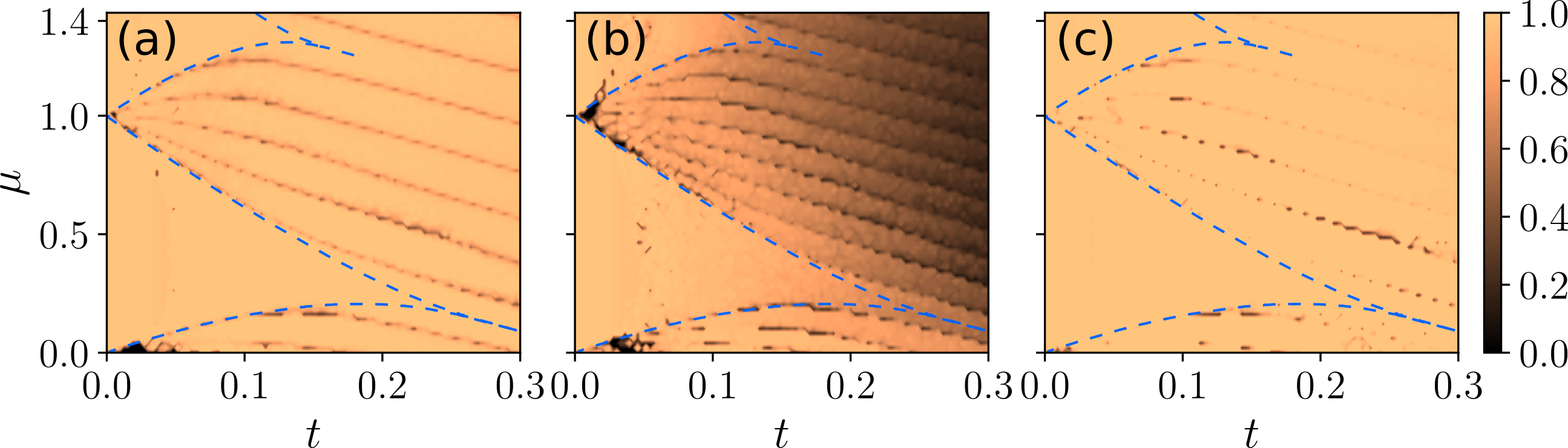}
    \caption{(Color online) Overlap $\abs*{\braket*{\psi_\text{exact}}{\tilde{\psi}_{\vb*{\theta}}}}^2$ between the exact and sampled RBM ground states for (a) 5 sites and 8 hidden neurons, (b) 8 sites and 11 hidden neurons and (c) 5 sites and 20 hidden neurons, for a maximum of 2048 states sampled from the Hilbert space. Dashed lines are the MI-SF boundaries as in \cref{fig:energy_convergence}.}
    \label{fig:overlapRBMsampled}
\end{figure}

Since the advantage of VMC over exact diagonalization comes for intractably large Hilbert spaces, it is not always possible to compute the probability amplitudes for all of the Fock states basis. In such a case, the RBM state can be built as
\begin{align}
    \ket*{\tilde{\psi}_{\vb*{\theta}}} = \frac{\sum_{\vb*{n}\in\mathcal{M}'}\psi_{\vb*{\theta}}(\vb*{n})\ket{\vb*{n}}}{\sqrt{\sum_{\vb*{n}\in\mathcal{M}'} |\psi_{\vb*{\theta}}(\vb*{n})|^2}},
\end{align}
where $\mathcal{M}'$ is a sample of Fock states sampled with the Metropolis-Hastings algorithm, with an acceptance probability of $\min\{1, p_{\text{GC}}(\vb*{n}_{i+1})/p_{\text{GC}}(\vb*{n}_{i})\}$, where $p_{\text{GC}}(\vb*{n}) = \mathcal{Z}^{-1}\exp(-\expval{\hat{H}}{\vb*{n}} )$~\footnote{If instead the sample is built from the distribution $\abs{\psi_{\vb*{n}}}^2$, then the acceptance probability becomes $\min\{1, |\psi_{\vb*{\theta}}(\vb*{n}_{i+1}) / \psi_{\vb*{\theta}}(\vb*{n}_i)|^2\}$ and the state build through \cref{eq:exactrbm} inherits the RBM state problems (data not shown).} is the probability associated with the grand canonical ensemble (note that the term $\mu\expval*{\hat{N}}$ has already been introduced in the Hamiltonian). This strategy was used to generate a sample $\mathcal{M}'$ of up to 2048 Fock states yielding a state $\ket*{\tilde{\psi}_{\vb*{\theta}}}$ for every point in the phase diagram. The overlap between the exact ground state $\ket{\psi_{\text{exact}}}$ and the sampled RBM ground state $\ket*{\tilde{\psi}_{\vb*{\theta}}}$ is shown in \cref{fig:overlapRBMsampled} for the three studied scenarios. 

Comparing \cref{fig:overlapRBMexact} with \cref{fig:overlapRBMsampled}, it is seen that the retrieved sampled state ``cleans'' the RBM state. In fact, the Fock states whose probability amplitudes were badly learned were removed, and only the Fock states relevant for the actual ground state were left. Despite this cleaning, the larger the Hilbert space, the more states have to be sampled to consider all relevant Fock states, as it is seen that 2048 Fock states are insufficient to capture the ground state in the case of 8 sites shown in \cref{fig:overlapRBMsampled}(b). Nonetheless, note that the low overlap in the arc from \cref{fig:overlapRBMexact}(b) almost completely disappears after the cleaning, cf. \cref{fig:overlapRBMsampled}(b). 

On the other hand, an important feature of \cref{fig:overlapRBMsampled}(b) is that the overlap of the sampled RBM state diminishes as $t$ gets larger. This happens because larger values of $t/U$ imply larger delocalization of the ground wavefunction, thus, involving more Fock states. Moreover, the Hilbert space size for 8 sites consists of $5^8 = 390625$ Fock states, which is why only sampling 2048 Fock states results in poor representations of the ground state, especially in the SF phase. Sampling more Fock states eventually reconstructs the exact ground state with very high overlap, except at the boundary between the MI and SF phases (data not shown).

\subsection{Order parameter}

\begin{figure}[!ht]
    \centering
    \includegraphics[width=\columnwidth]{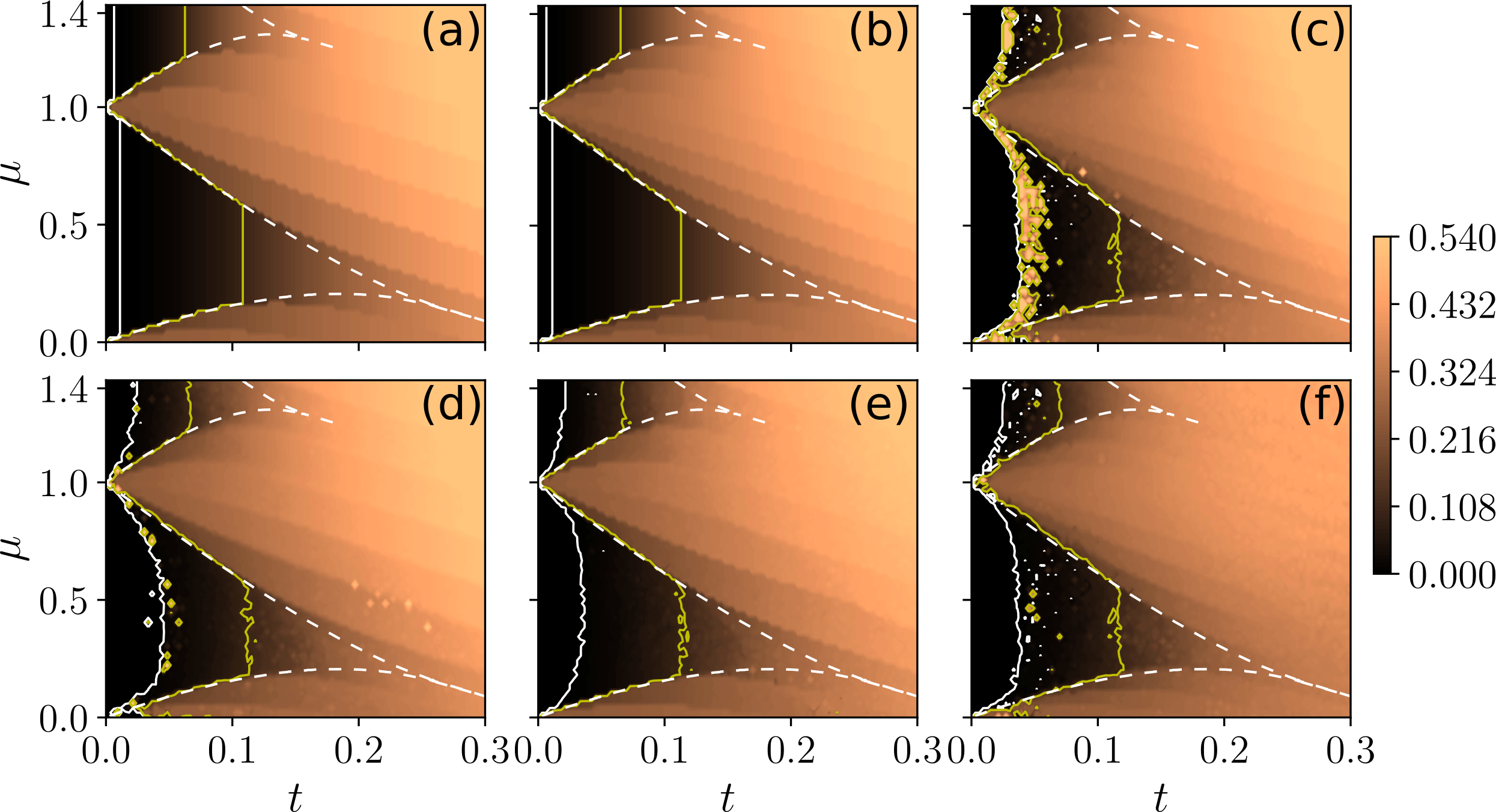}
    \caption{(Color online) Order parameter $\text{Var}(\hat{n}_1)$ for the ground state obtained through exact diagonalization for 5 sites (a), and 8 sites (b); for the sampled RBM ground state with 2048 Fock states for 5 sites with 8 hidden neurons (d) and 20 hidden neurons (e); and for the exact RBM ground state for 8 sites with 11 neurons (c) and the sampled RBM ground state (f). The white line is the 0.001 contour line, and the yellow one corresponds to the 0.01 contour line in all plots. Dashed lines are the MI-SF boundaries as in \cref{fig:energy_convergence}.}
    \label{fig:order_param}
\end{figure}

The phase diagram of the BH model can be reconstructed by measuring quantities in the sampled RBM ground state that exhibit the phase transition. As mentioned before, we chose the variance of the local number operator, in particular, of the first site. In \cref{fig:order_param}(a) and (b), the order parameter measured with the exact ground state is shown for 5 and 8 sites, respectively. It is seen that in the Mott insulator phase, the variance of the number of bosons in the first lattice site is near to 0, but not exactly 0 because of finite size effects. On the other hand, \cref{fig:order_param}(d) and (e) show the order parameter for the sampled RBM ground state with 2048 Fock states for the cases of 8 and 20 hidden neurons for 5 sites, which show excellent agreement with their exact counterpart \cref{fig:order_param}(a). However, at the boundary between the MI and SF phases, there are absolute errors that could be as high as 0.15, which come from the difficulty of learning the ground state near the phase transition boundary. In spite of these differences at the boundaries, it is clear that the learned RBM ground state mimics the re-entrance found in finite 1D chains~\citep{buonsante2007,park2004critical}. Finally, \cref{fig:order_param}(c) shows the order parameter for the exact RBM ground state for 8 sites and 11 hidden neurons. \Cref{fig:order_param}(f) also shows the order parameter for 8 sites and 11 hidden neurons but with a different sampling limit. Instead of fixing a number of Fock states to be sampled (2048 and 4096 were insufficient, data not shown), we build the sample $\mathcal{M}'$ by accepting states with the Metropolis-Hastings algorithm until there is a run of 400 consecutive state proposals that do not raise a new accepted state into $\mathcal{M}'$. In this case, the state is cleaned (note that the arc of badly learned order parameter almost completely disappears). A clear indicator of the number of Fock states to represent the ground state emerges: for very low values of $t$, within the first Mott lobe, only one sampled state ($\ket{1,1,1,1,1}$) is needed to reproduce the exact ground state; in the SF phase, up to 30000 states are sampled before hitting a 400 streak of no new accepted states into the sample.

\begin{figure}
    \centering
    \includegraphics[width=\columnwidth]{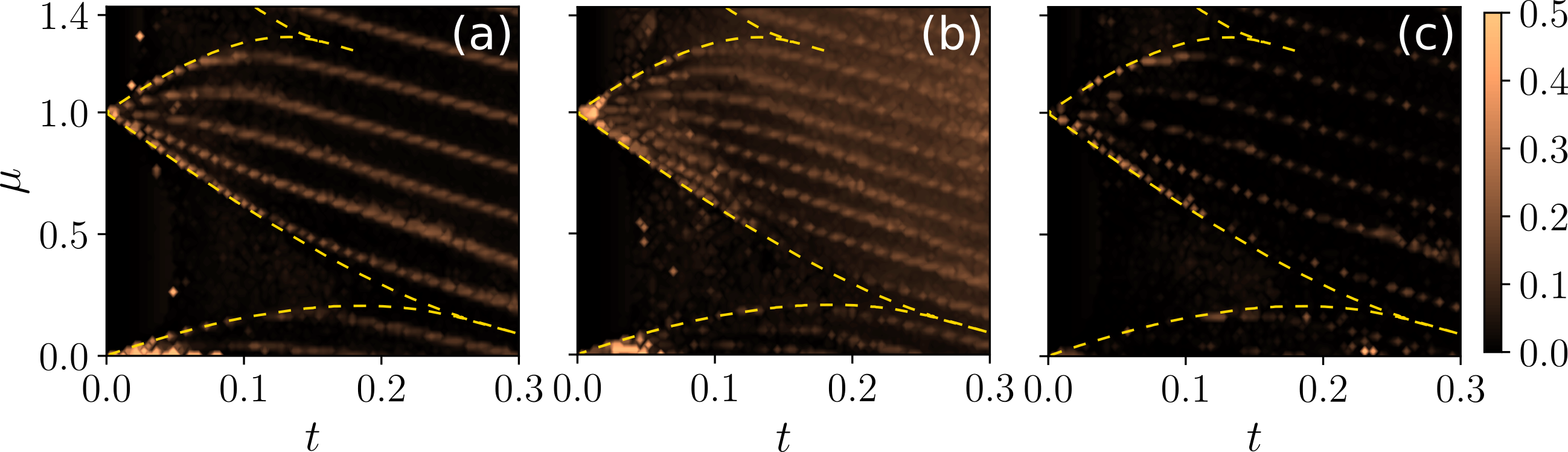}
    \caption{(Color online) Absolute error for the linear entropy at the first site of the chain between the exact ground state and the sampled RBM ground state for (a) 5 sites, 8 hidden neurons and 2048 Fock states, (b) 8 sites, 11 hidden neurons and 4096 Fock states, and (c) 5 sites, 20 hidden neurons, and 2048 Fock states. Dashed lines are the MI-SF boundaries as in \cref{fig:energy_convergence}.}
    \label{fig:order_param_entropy}
\end{figure}

Other quantities can be used as an order parameter, which more explicitly relate to quantum correlations such as entanglement. For instance, a partial trace carried out over all degrees of freedom except the first site yields a reduced density matrix $\rho_1(t,\mu)$, from which the linear entropy $S(t,\mu) = 1 - \Tr{\rho^2_1(t,\mu)}$ can be measured (the von Neumann entanglement entropy can also be used, e.g., \cite{Ejima2012}). \Cref{fig:order_param_entropy} shows the absolute errors between the exact and the sampled RBM ground states for the three studied scenarios, where the errors at the MI-SF boundary become large.

\subsection{Tomography}

\begin{figure*}
    \centering
    \includegraphics[width=\textwidth]{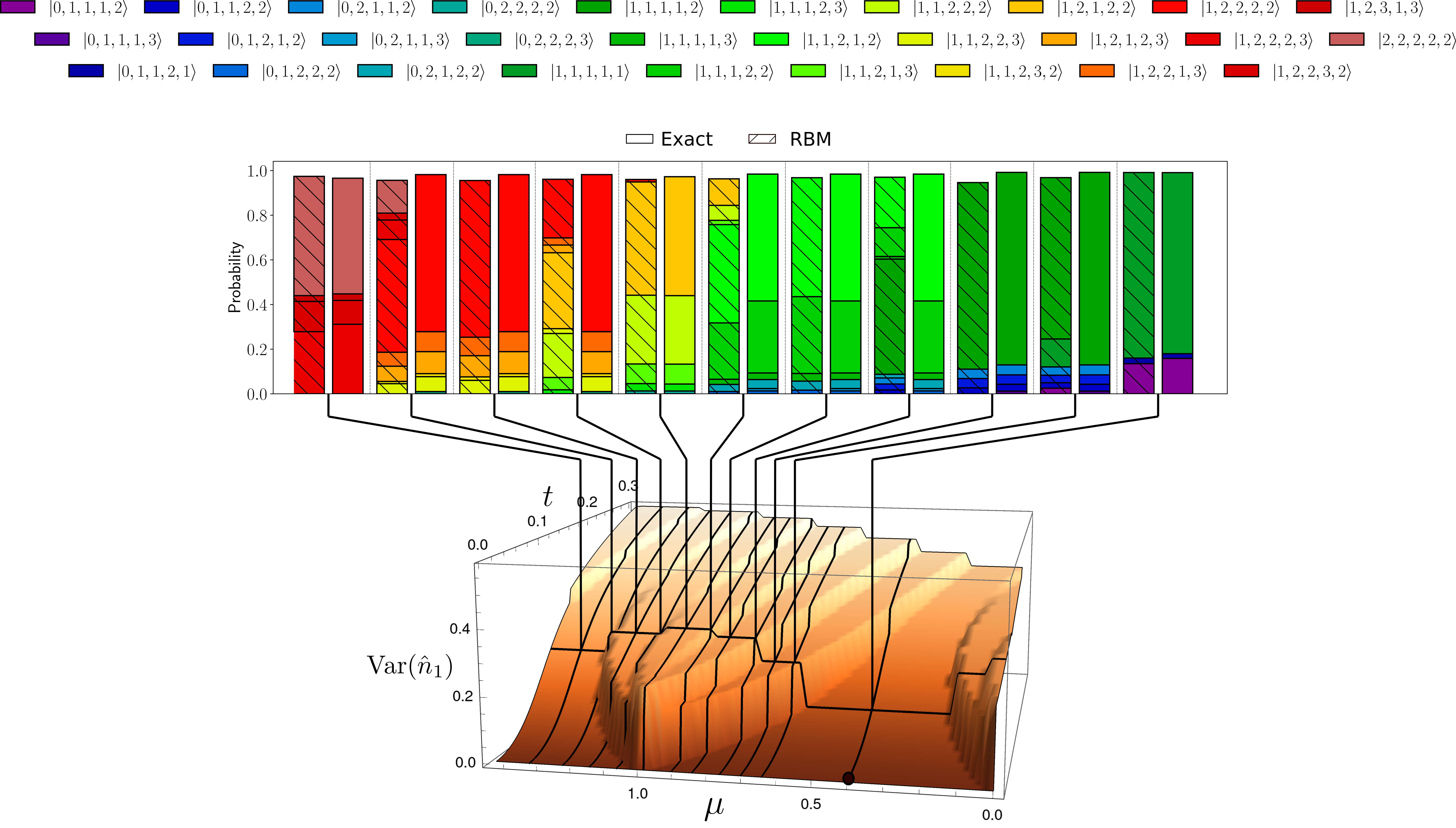}
    \caption{(Color online) Quantum tomography of the probability amplitude of Fock state manifold rungs for the ground state found through exact optimization and through VMC with an RBM wavefunction ansatz. In the lower part of the figure, a 3D plot of the order parameter for the exact ground state is shown for the case of 5 sites. Each bar color represents a manifold rung depicted through its canonical Fock state as the lexicographically smallest one (e.g. the rung $\{\ket{1,1,1,1,2}, \ket{1,1,1,2,1},\ldots,\ket{2,1,1,1,1}\}$ is represented by $\ket{1,1,1,1,2}$). A black dot in the middle of the first Mott lobe indicates that a tomography was made there as well.}
    \label{fig:tomos}
\end{figure*}

In all the studied scenarios, there are problems for learning the ground state at the MI-SF boundaries, as well as at the mini-plateaus boundaries within the SF phase. In order to understand the differences between the learned ground state and the one obtained through exact diagonalization, we performed a study of the composition of the ground states for 5 sites and 8 hidden neurons. For that reason, we examined the probability amplitudes of the ground state (both RBM learned and exact) at 11 different points in the $t$-$\mu$ space, fixing $t=0.1$, and with $\mu$ at the middle and border of each plateau in the MI and SF phases. We also examined the RBM and exact ground states for very small $t$ at the middle of the first Mott lobe, as indicated by the black dot in \cref{fig:tomos}, where the ground state for the RBM and for exact diagonalization was $\ket{1,1,1,1,1}$ with an associated probability amplitude of 99.99\%, as expected.

Note that the BH chain is invariant (up to a phase) under displacements and inversions, i.e. there are displacement and inversion operators that act as follows on Fock states: $\hat{T}\ket{n_1,\ldots,n_{N-1}, n_N} = e^{i\phi}\ket{n_N, n_1, \ldots, n_{N-1}}$ for displacement, and $\hat{I}\ket{n_1,n_2,\ldots,n_{N-1}, n_N} = e^{i\varphi}\ket{n_N, n_{N-1},\ldots,n_2,n_1}$ for inversion. Therefore, in order to perform a tomography, we must take into account that all Fock states that belong to the same rung defined by displacement and inversion operations are equivalent. Now, each Fock state can be brought to a canonical Fock state through a consecutive application of displacement and inversion operators onto the original Fock state. This canonical Fock state is selected as the lexicographically smallest one, after every possible application of displacement and inversion operators, as in~\citep{Choo2018}. In \cref{fig:tomos}, we show the probability amplitude distribution of the Fock state manifold rungs for the exact and the RBM ground states. More explicitly, the exact ground state is
\begin{align}
    \ket{\Psi} = \sum_{\vb*{n}} \Psi(\vb*{n})\ket{\vb*{n}} = \sum_{i}\sum_{\vb*{n}\in R_i}\Psi(\vb*{n})\ket{\vb*{n}},
\end{align}
where $i$ indexes the rungs, and $R_i$ is the $i$-th rung. The probability amplitude corresponding to a rung is, therefore, the sum of the probability amplitudes of all of its Fock states, and the bars from \cref{fig:tomos} show those rungs' probability amplitudes. Reading the plot from right to left, i.e. starting with the smallest value of $\mu$, we see that within the first Mott lobe, both the RBM and the exact ground states show very similar distributions over three rungs of the one-filling manifold: $\ket{1,1,1,1,1}$, $\ket{0,1,1,1,2}$ and $\ket{0,1,1,2,1}$. Increasing $\mu$ up to the first MI-SF boundary (slightly within the SF phase), we see that the probability amplitude distribution starts to differ between the exact and the RBM ground states, and the RBM struggles to identify if the ground state is now in an excitation manifold above the one-filling manifold, represented by the rung $\ket{1,1,1,1,2}$, or in the MI phase which is represented by the rung $\ket{1,1,1,1,1}$. Increasing again $\mu$ in order to be at the middle of the first plateau within the SF phase shows that both the RBM and the exact ground states have similar probability amplitude distributions, even though the RBM assigns a probability to other rungs (not shown in the plot because their contribution is less than 0.01). It is clear that this first plateau is mostly represented by the rung $\ket{1,1,1,1,2}$ which is in the 6-th excitation manifold. If we continue to increase $\mu$ we see that near the boundaries between the MI-SF phases and between the plateaus within the SF phase, the RBM and the exact ground states exhibit differences in the probability amplitude distributions. On the contrary, in the middle of those plateaus, the probability amplitude distributions from both RBM and exact ground states are very similar (which is also seen in \cref{fig:overlapRBMexact}(a)). Moreover, each time the tomography moves onto a new plateau (with a higher value of $\mu$), the excitation manifold increases by one, up to the 10-th excitation manifold, which corresponds to the two-filling manifold at the second Mott lobe, characterized by the rung $\ket{2,2,2,2,2}$ as well as the rungs $\ket{1,2,2,2,3}$ and $\ket{1,2,3,1,3}$. 

\section{Conclusions\label{sec:conclusion}}

In this work, we systematically tested the capabilities of VMC with a trial ground wavefunction given by an RBM on the one-dimensional Bose-Hubbard model. The motivation for the technique comes from the possibility of incorporating it into the toolbox of quantum physics to tackle theoretical problems that are difficult to study numerically due to the intractably large Hilbert spaces. Thus, it is first needed to intensively test the technique to reproduce known results, and it is also needed to theoretically explain why the technique works (this is a challenging open question which involves the question of why neural networks work well). Only if the community identifies the strengths and weaknesses of the technique is it possible to use it to explore problems of interest that involve vast Hilbert spaces. Regarding the model under study, we repeatedly found differences between the exact ground state found through the exact diagonalization of the BH Hamiltonian and the learned ground state. We did so in one-dimensional chains of 5 and 8 sites. In order to better learn the ground state, the variational trial wavefunction was enriched in the case of 5 sites by increasing the number of hidden units in the RBM's hidden layer. Although results improved near the MI-SF boundary, other regions of the BH parameter space suffered from badly learned ground states. For that reason, we proposed a sampling technique that cleans the ground state, getting rid of Fock states not relevant to the actual ground state. 

In most of the BH model parameter space, it was found that the VMC-RBM method yielded good results in computing the energy (with low errors and clear signs of convergence), and also was able to compute other observables which explicitly involve quantities related to quantum correlations, such as the linear entropy of one of the chain's sites. Moreover, we reconstructed with excellent accuracy the phase diagram of the BH model using the local occupation variance as an order parameter for most of the parameter space. Furthermore, we also carried out quantum state tomographies to understand the composition of the ground states at the interfaces between the MI-SF phases and also within the excitation manifold plateaus formed for finite chains in the SF phase, which revealed significant differences between the Fock states distributions of the exact ground state and the learned ground state.

Accordingly, we must raise the attention that the VMC technique is consistently challenged near the quantum phase transition of the BH model. We do not doubt that the high-expressivity of neural quantum states is able to improve those results under more extensive sampling (or better sampling schemes), or more complex variational trial wavefunctions apart from RBMs; however, we report that some questions have to be answered with precision before confidently using VMC as a method to accurately explore the physics of a many-body problem without the support of any other numerical method. In the first place, we observed that not every point of the phase diagram requires the same amount of computational work to learn the ground state. Most notably, the ground state in the MI phase is mostly explained by only one Fock state, whereas the number rapidly grows for ground states in the SF phase. Therefore, automated ways of stopping sampling within the sampling steps should be taken into consideration. In particular, when using the Metropolis-Hastings algorithm, we found that if a Markov chain of a certain length was formed with no new sampled states, we could stop sampling, yielding a high-quality ground state. Secondly, the VMC technique showed badly learned ground states along arcs formed within the Mott lobes. Why these form remains unanswered, as they are not related to the structure of the MI-SF boundaries. However, after we clean the learned ground state, these arcs almost completely disappear. Thus, it might be the case that they are related to noisy probability from Fock states that are not relevant to the ground state. Finally, concerning the first question, the number of optimization steps required to learn the ground state has to be better understood, as it is only clear when the energy converges, but not the state~\footnote{In this regard, NetKet team has included the Gelman-Rubin statistics~\citep{vats2018revisiting}}.

Additionally, an interesting discussion arises when paying close attention to the sampling technique of the VMC. Netket's Local Metropolis sampler has low acceptance probability near the MI-SF transition for low values of $t$ and finite small chains such as the ones considered in our work, which complicates learning the state. However, the difference between the exact state and the learned state is constant throughout manifold excitation boundaries regardless of the $t$ value, and also regardless of being at the MI-SF transition, or within the SF transition.
\begin{acknowledgments}
We want to thank Andrés Urquijo and J. P. Restrepo-Cuartas for useful discussions at the early stage of this study.
\end{acknowledgments}
\bibliography{69725.bib}% Produces the bibliography via BibTeX.

\end{document}